\documentclass[journal=langd5,manuscript=article]{achemso}
\pdfoutput=1
\usepackage[version=3]{mhchem} % Formula subscripts using \ce{}
\usepackage{amsmath,amssymb}
\usepackage{graphicx}% Include figure files
\usepackage{color}% Include colors for document elements
\usepackage{dcolumn}% Align table columns on decimal point
\usepackage[separate-uncertainty,multi-part-units=single]{siunitx} % proper units usage
\usepackage{makecell}
\usepackage[font=normal]{subcaption}
\usepackage{booktabs}
\usepackage{multirow}
\DeclareSIUnit{\atm}{atm}
\DeclareSIUnit{\calorie}{cal}
\DeclareSIUnit{\debye}{D}
\DeclareSIUnit{\rydberg}{Ry}

\author{Henrique M. Cezar}
\affiliation[Universidade de Sao Paulo]{Universidade de Sao Paulo, Instituto de Fisica, Rua do Matao 1371, Sao Paulo, SP, 05508-090, Brazil}
\alsoaffiliation[Hylleraas Centre]{Hylleraas Centre for Quantum Molecular Sciences and Department of Chemistry, University of Oslo, PO Box 1033 Blindern, 0315 Oslo, Norway}
\email{h.m.cezar@kjemi.uio.no}
\author{Caetano R. Miranda}
\affiliation[Universidade de Sao Paulo]{Universidade de Sao Paulo, Instituto de Fisica, Rua do Matao 1371, Sao Paulo, SP, 05508-090, Brazil}
\email{crmiranda@usp.br}

\title{Water adsorption in ultrathin silica nanotubes}

\begin{document}

  % \linenumbers

  \section{Abstract}
  \label{abstract}

  Silica (\ce{SiO2}) nanotubes (NTs) are used in a wide range of applications that go from sensors to nanofluidics.
  Currently, these NTs can be grown with diameters as small as \SI{3}{\nano\meter}, with walls \SI{1.5}{\nano\meter} thick.
  Recent experimental advances combined with first-principles calculations suggest that silica NTs could be obtained from a single silica sheet.
  In this work, we explore the water adsorption in such ultrathin silica NTs using molecular simulation and first-principles calculations.
  Combining molecular dynamics and density functional theory calculations we obtain putative structures for NTs formed by \num{10}, \num{12} and \num{15}-membered \ce{SiO2} rings.
  Water adsorption isotherms for these NTs are obtained using Grand Canonical Monte Carlo simulations.
  Computing the accessible cross-section area ($A_\text{free}$) for the NTs, we were able to understand how this property correlates with condensation pressures.
  We found that $A_\text{free}$ does not necessarily grow with the NT size, and that the higher the confinement (smaller $A_\text{free}$), the larger the condensation pressure.

  \section{Introduction}
  \label{sec:intro}

  Silica (\ce{SiO2}) is an inorganic material that is often used in aqueous solution for different kinds of applications in adsorption,\cite{Taheri2019} catalysis,\cite{Bivona2019} separation,\cite{Ding2020} sensors,\cite{ALTALHI2021} energy,\cite{Yan2020,Zhang2017} thermal isolation,\cite{Syme2020} nanofluidics\cite{Zhu2017} among other fields.
  The great thermal stability, combined with a variety of stable structures, control over hydrophobicity, and the possibility of functionalization with organic groups are some of the reasons why silica is employed in such a wide range of applications.
  In particular, for nanofluidics, the possibility of synthesizing nanotubes (NTs) with a fine control of its diameters control over the hydrophobicity through functionalization, makes the silica NTs an excellent platform for nanofluidic studies.

  The first report of the synthesis of silica NTs is from Nakamura \textit{et al}.\cite{Nakamura1995}
  Ever since, amorphous silica NTs have been synthesized and characterized with various shapes and inner diameters ranging from \SI{3}{\nano\meter}\cite{Wang2015,Yuan2013,Lin2010} to hundreds of nanometers.\cite{Yang2011,Jung2010,GarciaCalzon2012,Farid2017,Farid2020}
  The wall thickness of these NTs is also a few nanometers.
  A silica double-walled NT has also been synthesized,\cite{Pouget2007} with an inner diameter of about \SI{15}{\nano\meter} and an ultra-thin wall thickness of just \SI{1.4}{\nano\meter}.
  More recently, efforts have been done to change the NT wall properties by combining materials such as \ce{Au} nanoparticles,\cite{Kong2020}, \ce{Sn} atoms\cite{Bivona2019}, or organic ligands\cite{Li2018,Mandal2018,Oredipe2019} for use in different types of applications.

  Another class of silica materials that have been synthesized are the supported silica monolayers and silica bilayers.\cite{Buchner2017}
  These ultrathin films are typically grown supported on metal substrates. 
  In the case of bilayers, the film is quasi-freestanding and may have applications in the separation of gases.\cite{Yao2017}
  Silica monolayers, on the other hand, can be synthesized supported in metals with high oxygen adsorption energy.\cite{Shaikhutdinov2013}
  These monolayers are composed of corner-sharing \ce{SiO4} tetrahedrons and have a \ce{SiO_{2.5}} composition.

  Based on density-functional theory calculations, it has been proposed that the formation of ultrathin silica NTs formed from a single monolayer\cite{Zhao2006,Zhao2007,Fang2015} (analogous to carbon nanotubes) or double layers\cite{Zhou2014} may be possible.
  The single-layer formation energies are negative\cite{Fang2015}, and strain energies are smaller than the already synthesized silica fiber.\cite{Zhao2006,Zhao2007}
  For double-layered NTs, the energy required to roll a double-layered NT is higher but comparable to the energy used to roll a carbon nanotube.\cite{Zhou2014}
  The ends of the silica NTs should be metastable in the presence of water,\cite{DeLeeuw2003} making the synthesis challenging.

  However, advances have been made in the last few years.
  Recently, Wang and coauthors\cite{Wang2021} reported a method for the electrochemical conversion of silica nanoparticles into silicon nanotubes.
  During the electrochemical process, \ce{SiO_x} ($\num{0} < x < \num{2}$) sheets are exfoliated, and after longer times the layers are crimped and form the silicon NTs.
  Before the work of Wang et al., silicon NTs were usually obtained with the aid of templates and chemical vapor deposition.

  In this work, we investigate the structure and water adsorption in single-layered ultrathin silica NTs.
  The interface of such NTs with water has been previously studied but not in a systematic manner.\cite{He2008}
  We combine classical force field modeling with density functional theory (DFT) calculations to find the structures of \num{10}, \num{12} and \num{15}-membered-rings NTs.
  Using molecular dynamics (MD) simulations, we perform annealing of such NTs using different force fields and use DFT to optimize these structures.
  The NTs are characterized concerning their accessible cross-section area ($A_\text{free}$), which is related to the accessible pore volume.

  To select the best general-purpose silica force field for water adsorption, we performed Grand Canonical Monte Carlo (GCMC) simulations of water in the silicalite-1, a pure silica zeolite with \num{10}-membered-rings channels.
  Comparing the adsorption isotherms obtained using the CLAYFF,\cite{Cygan2004} Cruz-Chu\cite{Cruz-Chu2006}, and Interface\cite{Emami2014} force-fields we selected the Interface force-field for the NTs simulations, as it qualitatively provided the best description of the experimental isotherms.
  
  Finally, we perform water adsorption studies in these NTs using GCMC simulations.
  We started with GCMC/MD cycles, loading the NTs with water at high pressure with GCMC followed by a short MD that allowed the NTs to relax and accommodate the water molecules.
  The GCMC/MD cycles were repeated until the convergence of the number of adsorbed molecules.
  Starting from the converged configuration of the NTs, we used GCMC simulations to calculate the adsorption isotherm of \num{5} different NTs of different shapes and sizes.

  \section{Methodology}
  \label{sec:methodology}

  \subsection{Structural modeling}
  We used a \ce{SiO2} unit in a square lattice to build monolayers with a \num{4}-membered \ce{SiO2} ring (4MR).
  The 4MR monolayer was chosen based on the Zhao \textit{et al.} work\cite{Zhao2007} which found that NTs formed from this monolayer had lower strain energy, and the work of Zhang \textit{et al.} which shown that these NTs are energetically viable and thermally stable.\cite{Zhang2006}
  The unit cell was replicated to form the silica monolayer which later was rolled into the nanotubes.
  Three diameters were considered: of about \SI{11}{\angstrom}, \SI{12}{\angstrom} and \SI{15}{\angstrom}, corresponding to the (\num{10},\num{0}), (\num{12},\num{0}) and (\num{15},\num{0}) perfectly cylindrical NTs.

  This initial structure was then submitted to annealing cycles as described below in the MD section.
  We selected up to \num{6} structures obtained with the annealing for further optimization with DFT.
  The length of the nanotube was of 2 \ce{SiO2} units along the $z$ axis in the case of the annealing and DFT calculations and about \SI{15}{\nano\meter} for the GCMC simulations.
  For the silicalite-1 simulations we used a $\num{2}\times\num{2}\times\num{2}$ structure replicated from the MFI IZA-SC\cite{IZA-SC} unit cell.

  \subsection{Force fields}
  We investigate the stability of the ultrathin NTs and the water adsorption within three different classical force field models.
  There are different force fields suitable to describe the silica-water interactions available in the literature, such as the Cruz-Chu,\cite{Cruz-Chu2006} Interface,\cite{Emami2014} CLAYFF,\cite{Cygan2004} Lopes\cite{Lopes2006} and Cole\cite{Cole2007} force fields.
  In this work, we compare the Cruz-Chu, Interface, and CLAYFF for the description of the structure and water adsorption properties.
  For the adsorption studies, the structures are considered completely rigid and interact with water only through the Lennard-Jones and Coulomb potentials.
  During the search for the energy minimums of each potential (described below in the MD simulations) all the suitable bonded interactions are considered.

  The water was simulated using the SPC/E\cite{Berendsen1987} model, as it is compatible with the selected force fields.
  However, we have also considered the Cruz-Chu potential with the TIP3P\cite{Jorgensen1983} water model for comparisons.
  Tables containing the parameters for each force field and water model are provided in the Supporting Information.

  \subsection{Molecular dynamics simulations}
  We use molecular dynamics (MD) to simulate the annealing of NTs and generate candidate structures.
  At this stage, no water molecules were considered.
  We tested different protocols, and employed at least \num{2} annealing cycles.
  First, we heated the NT from \SI{10}{\kelvin} to \SI{10000}{\kelvin}, and followed with a \SI{300}{\pico\second} MD on the final temperature.
  Later, we cooled from \SI{10000}{\kelvin} to \SI{1000}{\kelvin}, performed a short \SI{100}{\pico\second} MD at \SI{1000}{\kelvin}, and finally performed another cooling from \SI{1000}{\kelvin} to \SI{10}{\kelvin}, ending with a \SI{50}{\pico\second} MD at this temperature.
  The final structure was minimized with the conjugate gradient algorithm until the energy convergence with a criterion of \num{e-20} \SI[per-mode=symbol]{}{\kilo\calorie\per\mol}.
  
  Molecular dynamics was also used after adsorption in cycles with GCMC to relax the NTs to better accommodate the water molecules.
  In this case, the NTs loaded with water from the GCMC simulations were used as the initial configuration of MD. 
  These simulations were performed for \SI{100}{\pico\second}, with temperature starting at \SI{300}{\kelvin} and being lowered to \SI{10}{\kelvin} along the dynamics.

  The equations of motion were integrated with the velocity verlet algorithm with a \SI{1}{\femto\second} timestep using the Nose-Hoover thermostat with a damping factor of \num{100} timesteps, removing the linear and angular momentum to avoid the flying ice cube.
  In the water simulations, the SHAKE algorithm was used to keep the water molecules rigid.
  All the MD simulations were performed using LAMMPS.\cite{Plimpton1995} 

  \subsection{Density functional theory calculations}
  Our structural and energetic results with each of the force field models were compared with first-principles density functional theory (DFT) calculations.
  The most stable configurations found by the annealing performed with the classical force fields were submitted to further DFT optimizations using the PBE\cite{Perdew1996} exchange-correlation functional.
  Since water molecules were not considered at this stage, the use of a functional with improved van der Waals description would not bring large differences on the structures.
  We performed a full relaxation of the simulation cell composed of an NT with a length of two rings in the axial direction.
  Considering rings with \num{10}, \num{12} and \num{15} \ce{SiO2} units, the NTs had a total of \num{60}, \num{72} and \num{90} atoms, respectively.
  Projector augmented wave pseudopotentials were employed, and a self-consistent criterion of \num{e-6} was used for the energies.
  As usual, the energy convergence with the plane-wave energy cutoff and the number of \textit{k} points was tested, and a cutoff of \SI{32}{\rydberg} and a Monkhorst-Pack generated \num{1} $\times$ \num{1} $\times$ \num{4} \textit{k}-point grid with a half step offset were employed.
  The calculations were performed using Quantum Espresso.\cite{Giannozzi2009}

  \subsection{Grand-canonical Monte Carlo simulations}
  We investigate the adsorption of water in the liquid phase at \SI{300}{\kelvin} and different pressures in the NTs employing GCMC simulations.
  First, we performed pure liquid water simulations in the grand-canonical ensemble for different values of the chemical potential $\mu$.
  % We associate the pressure to each $\mu$ using the Desbiens \textit{et al.} protocol.\cite{Desbiens2005}
  The GCMC simulations were run for at least \num{e9} steps in cubic boxes of \SI{30}{\angstrom} to determine the average density for each chemical potential.
  We used a configurational bias Monte Carlo scheme with \num{16} trial insertion positions in each step to improve the insertion acceptance rates.
  Translation, rotation, insertion, and deletion moves were attempted randomly with equal probability.
  % Finally, simulations in the isotherm-isobaric ($NPT$) ensemble were used to associate the densities to pressures, determining $\mu(P)$.
  % In these simulations, volume change attempts were performed with a \SI{0.5}{\percent} probability, while translation and rotation moves were attempted with \SI{49.75}{\percent} each.

  The water adsorption isotherms of the NTs were performed with restricted insertions and were run for at least \num{5e6} MC steps, until the convergence of the energy and number of adsorbed molecules.
  Water molecules were inserted in a cylinder with the radius of the NT, also using configurational bias insertions to improve the acceptance rates.
  The translation, rotation, insertion, and deletion moves were also attempted with \SI{25}{\percent} each.
  All the MC simulations used a Lennard-Jones cutoff of \SI{12}{\angstrom} and used the Ewald summation with a cutoff of \SI{12}{\angstrom} in the real space and accuracy of \num{e-5} for the electrostatics.
  Simulations for silicalite-1 followed a similar protocol, but without restricted insertions and were run for \num{1.5e8} steps.
  All the Monte Carlo simulations were performed using Cassandra.\cite{Shah2017}

  \section{Results and discussion}
  \label{sec:results_disc}

  \subsection{Putative global minima of the NTs}

\begin{figure*}[t]
    \begin{center}
      \includegraphics[width=\textwidth]{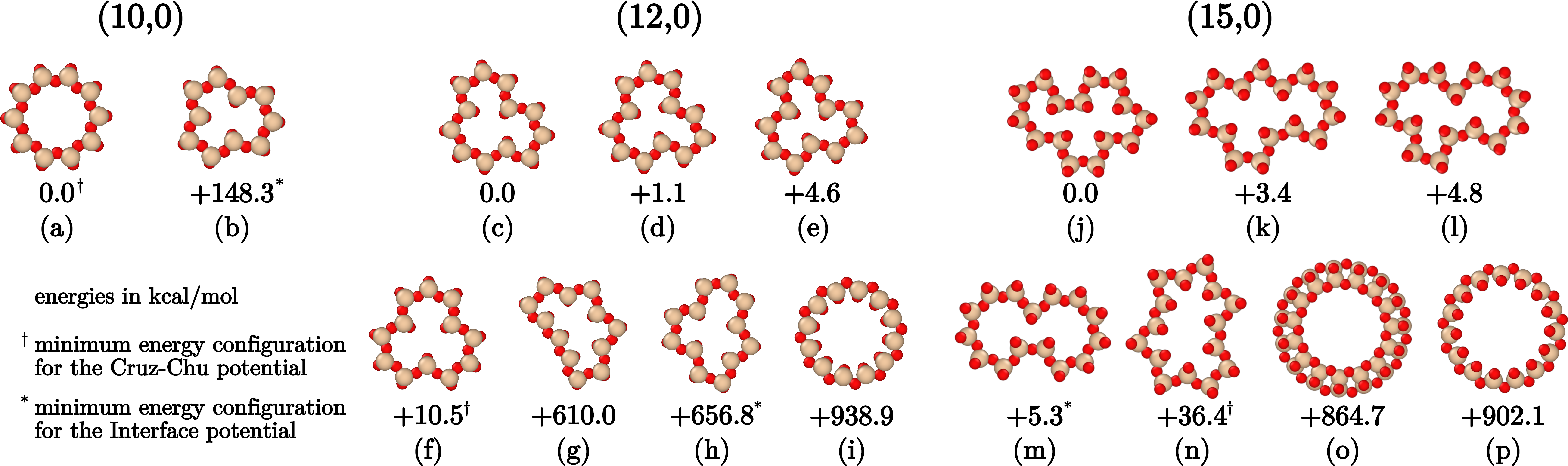}
    \end{center}
    \caption{\label{fig:structure} Putative minimum energy structures of the silica NTs obtained with DFT. Energies are in \SI[per-mode=symbol]{}{\kilo\calorie\per\mol} and are displayed as the energy increase from the minimum energy for each size. The symbols represent the minimum energy structure for the Cruz-Chu ($^\dagger$) and Interface ($^*$) force fields for each size.}
  \end{figure*}

  To search the minimum energy configurations of the silica NTs we employ annealing cycles considering the Cruz-Chu and Interface force fields.
  The two force fields led to different minima which were further optimized using DFT.
  In total two (\num{10},\num{0}) NTs, seven (\num{12},\num{0}) NTs and seven (\num{15},\num{0}) were considered for the DFT optimizations, including the perfectly cylindrical NTs.
  The DFT energies and relative stabilities are shown in Figure~\ref{fig:structure}, where the minimum energy configurations of the Cruz-Chu and Interface force fields are highlighted.
  Overall, we observe that the putative global minima of the force fields are not the minimum energy structures given by DFT.
  For example, the lowest energy configuration for the (\num{12},\num{0}) NT with the Interface force field is given by (h), which is \SI[per-mode=symbol]{656.8}{\kilo\calorie\per\mol} higher than the minimum energy with DFT (configuration (c)).
  However, considering this same force field, the energy of (c) is just \SI[per-mode=symbol]{6.1}{\kilo\calorie\per\mol} higher than the energy of (h).

  We illustrate the differences between the minimum energy structures obtained with DFT and the ones used as the starting configurations --- minimum configuration from the force fields --- by the RMSD of the structures.
  The RMSDs were calculated using the Kabsch algorithm,\cite{Kabsch1976,Charnley2020} which gives the rotation of coordinates that minimize the RMSD.
  In Table~\ref{tab:area_rmsd} we show these RMSDs for each structure from (a) to (p) as illustrated in Figure~\ref{fig:structure}.
  The RMSDs are all smaller than \SI{1.0}{\angstrom}, except for structure (o), a high-energy configuration which had an RMSD of \SI{1.16}{\angstrom}.
  Such small values for the RMSD indicate that the structures optimized with DFT are similar to the force field configurations, which is indeed observed by visual inspection of the configurations.

\begin{table}[ht]
\caption{\label{tab:area_rmsd}Accessible cross-section area for the DFT geometry ($A^\text{DFT}_\text{free}$) and after the NT was filled with water ($A^\text{filled}_\text{free}$), and RMSD between the DFT and force field optimized structures for all the studied structures shown in Figure~\ref{fig:structure}. The filled area was calculated after removing the water molecules from the NT.}
\begin{tabular}{@{}ccccc@{}}
\toprule
                        & Structure & \begin{tabular}[c]{@{}c@{}}$A^\text{DFT}_\text{free}$\\ (\AA$^2$)\end{tabular} & \begin{tabular}[c]{@{}c@{}}$A^\text{filled}_\text{free}$\\ (\AA$^2$)\end{tabular} & \begin{tabular}[c]{@{}c@{}}RMSD\\ (\AA)\end{tabular} \\ \midrule
\multirow{2}{*}{(10,0)} & (a)         & 11.79 & 19.36                                                       & 0.26                                              \\
                        & (b)         & 0.01 &                                                     & 0.49                                              \\ \midrule
\multirow{7}{*}{(12,0)} & (c)         & 2.22 & 9.1                                                         & 0.60                                              \\
                        & (d)         & 0.11 &                                                      & 0.50                                              \\
                        & (e)         & 0  &                                                                & 0.57                                              \\
                        & (f)         & 1.92 &                                                       & 0.63                                              \\
                        & (g)         & 0.24 &                                                           & 0.48                                              \\
                        & (h)         & 1.46 & 0.56                                                         & 0.46                                              \\
                        & (i)         & 24.48 &                                                       & 0.52                                              \\ \midrule
\multirow{7}{*}{(15,0)} & (j)         & 0 &                                                                 & 0.63                                              \\
                        & (k)         & 1.65 &10.26                                                         & 0.72                                              \\
                        & (l)         & 3.66 &                                                         & 0.89                                             \\
                        & (m)         & 7.59 & 12.98                                                         & 0.73                                              \\
                        & (n)         & 6.21 &                                                           & 0.33                                              \\
                        & (o)         & 44.18 &                                                       & 1.16                                              \\
                        & (p)         & 41.14 &                                                        & 0.55                                              \\ \bottomrule 
\end{tabular}
\end{table}

  Among the different structures for a same size shown in Figure~\ref{fig:structure}, there are differences between the accessible volumes for adsorption.
  The accessible volume for each nanotube was calculated using a Monte Carlo integration scheme.\cite{Herrera2010}
  We randomly inserted \ce{Ne} atoms one at a time inside the NT in a volume $V_\text{out}$, and evaluated the interaction energy between the probe and the NT.
  A point inside the NT was considered as belonging to the accessible volume if the interaction energy was negative, and outside the accessible volume otherwise.
  The accessible volume is then obtained by
  \begin{equation}
    \label{eq:vfree}
    V_\text{free} = \frac{N_\text{in}}{N_\text{trial}}V_\text{out}, 
  \end{equation}
  with $N_\text{in}$ being the number of points counted as inside the accessible volume and $N_\text{trial} = \num{e6}$ is the number of trial insertions.
  Such a scheme was implemented in a LAMMPS script.
  
  In Table~\ref{tab:area_rmsd} we show the accessible cross-section area
  \begin{equation}
    \label{eq:afree}
    A_\text{free} = \frac{V_\text{free}}{L}
  \end{equation}
  where $L \approx \SI{150}{\angstrom}$ is the NT length.
  Two values of $A_\text{free}$ are reported, $A^\text{DFT}_\text{free}$ and $A^\text{filled}_\text{free}$.
  In this section, we focus on $A^\text{DFT}_\text{free}$, which are the accessible cross-section area of the optimized DFT structures, leaving the discussion of $A^\text{filled}_\text{free}$ for the adsorption section.

  We see that $A^\text{DFT}_\text{free}$ can change drastically depending on the configuration, even for the same size of NT.
  For example, for structures (b), (e) and (j) $A^\text{DFT}_\text{free} = \num{0}$, indicating that even for high pressures, water is unlikely to be adsorbed in the NT.
  The most stable configuration for the (\num{12},\num{0}) and  (\num{15},\num{0}) NTs all show a smaller $A_\text{free}$ than (a), the most stable (\num{10},\num{0}) configuration.
  Therefore, it is expected that the confinement effects will be greater for the (\num{12},\num{0}) and (\num{15},\num{0}) NTs.
  The complex potential energy surfaces of these NTs make challenging identifying the most stable configurations.
  Also, since the silica NTs do not have a cylindrical shape for every NT size, we show in the next section that even $A_\text{free}$ can be misleading in determining how confined a fluid can be inside the NT.
  
  \subsection{Water adsorption isotherms}

  We start validating the force fields for water adsorption in pure silica nanoporous materials.
  We chose the silicalite-1 zeolite as the platform for the validation as it there are experimental adsorption isotherms for this zeolite with varying degrees of local defects.\cite{Trzpit2007}
  The defects in silicalite-1 introduce silanol groups at the surface, which makes the material less hydrophobic.
  Therefore, we compare our results with the ones obtained for silicalite-1 synthesized from a fluoride route, which is known to contain fewer defects than using other methods.\cite{caullet2005}

\begin{figure}[t]
    \begin{center}
      \includegraphics[width=0.45\textwidth]{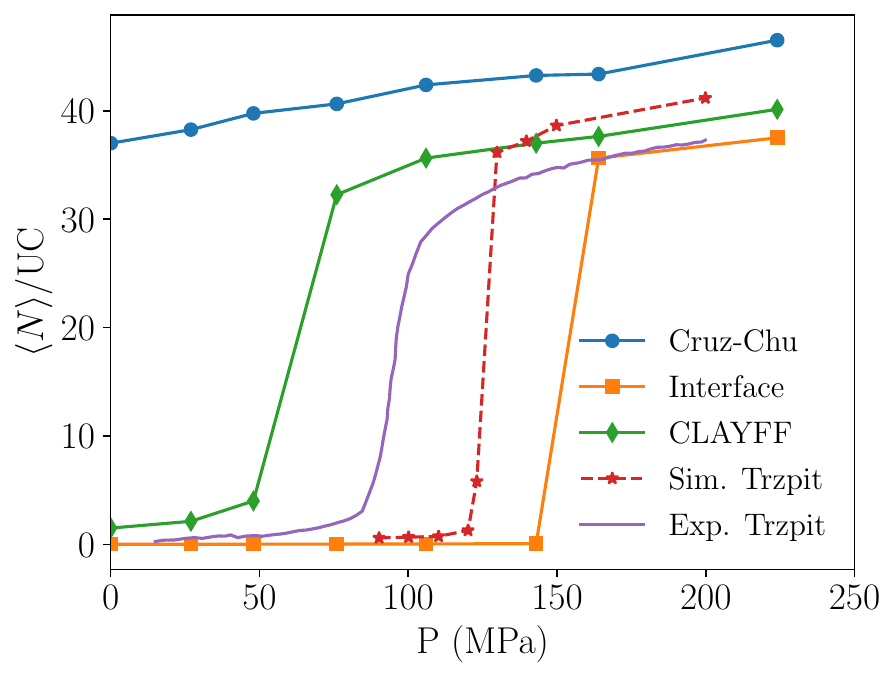}
    \end{center}
    \caption{\label{fig:silicalite_ads} Water adsorption isotherms for the silicalite-1 MFI zeolite. We show our results using three different force fields generally applied for silica, namely, the Cruz-Chu, CLAYFF and Interface force fields. We also display the experimental and simulation results of Trzpit et al.\cite{Trzpit2007}, with the experimental results corresponding to the zeolite with the least amount of local defects.}
  \end{figure}

  In Figure~\ref{fig:silicalite_ads} we show the experimental and simulated water adsorption isotherms from Trzpit et al.\cite{Trzpit2007} and our adsorption isotherms using three different general force fields for silica/water interfaces.
  The experimental results show that the zeolite starts filling at about \SI{80}{\mega\pascal}.
  A better match between the experimental and simulated isotherms can be achieved by adjusting the 
  \ce{Si} and \ce{O} atomic charges.\cite{Trzpit2007,Desbiens2005}
  However, this was not our goal.
  Here, we wanted to compare the three force fields, which can be used for different kinds of silica/water interfaces, so we could select a force field for the NT simulations.

  It is known that that the smaller the amount of silanol surface defects, the more the condensation transition is shifted towards higher pressures, and steeper curves when it fills\cite{Trzpit2007,Rother2020}.
  For Figure~\ref{fig:silicalite_ads}, we selected the isotherms with the smallest amount of defects as reported Trzpit and coauthors\cite{Trzpit2007}.
  Bearing this in mind, since the simulated zeolite was perfectly crystalline, we conclude that the Interface force field is the more suitable to describe the water/silica interactions for these pore sizes.
  We see that the Cruz-Chu force field overestimates the amount of adsorbed water at lower pressures, indicating that the capillarity pressure with this force field results in condensed water even at small pressures.
  Moreover, despite the reasonable description of the isotherm by CLAYFF, the force field was not able to hold the NTs together during MD simulations, limiting its use to rigid framework studies.

  Based on the $A^\text{DFT}_\text{free}$ and energy stability of the NTs shown in Figure~\ref{fig:structure} we selected \num{5} NTs to perform the adsorption studies, namely, (a), (c), (h), (k) and (m).
  We start by performing GCMC/MD cycles to relax the NT configurations in the presence of water.
  The DFT optimized configuration for each of these NTs was used as the initial configuration for a GCMC simulation at \SI{1080}{\mega\pascal}.
  After the number of molecules inside the NT converged, we performed an energy minimization and short MD (\SI{100}{\pico\second}) to enable the NTs to relax and arrange the atoms' positions to accommodate the water molecules.
  This process was repeated until the number of molecules inside the NT did not vary among two GCMC runs, which was about \num{2} or \num{3} GCMC/MD cycles.

\begin{figure*}[t]
    \begin{center}
      \includegraphics[width=\textwidth]{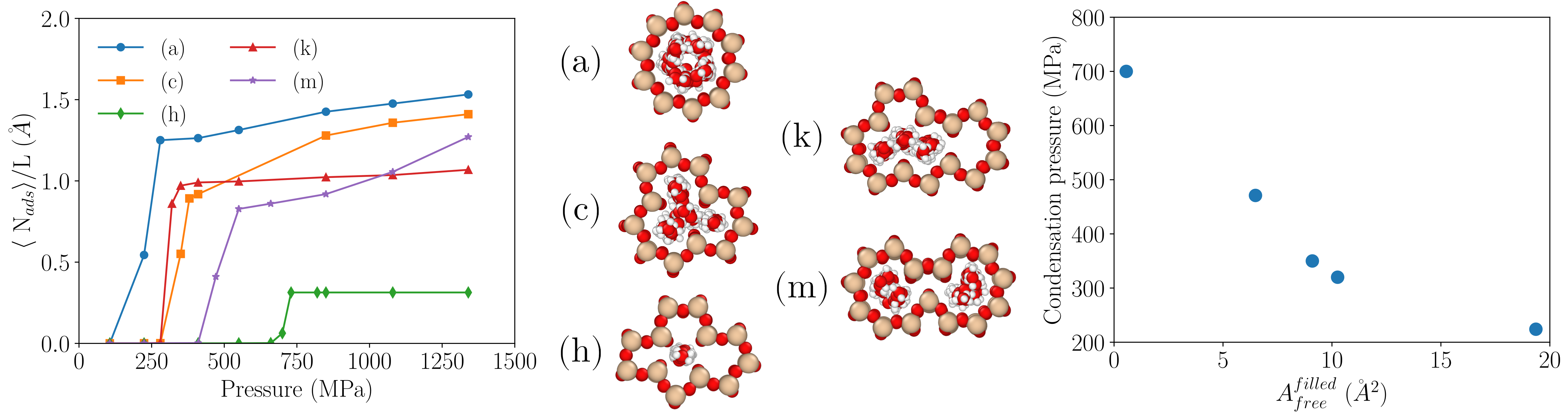}
    \end{center}
    \caption{\label{fig:full_isotherms} Water adsorption isotherms for different silica NTs (left). On the center, we show a snapshot for each filled NT, correspondent to the \SI{1340}{\mega\pascal} pressure. On the right panel, we show  a plot of the condensation pressures in function of the cross-section accessible area $A^\text{filled}_\text{free}$. We considered the condensation pressure as the pressure of the point with the highest slope. Uncertainties were slightly larger than the size of the points and were omitted for a cleaner plot.}
  \end{figure*}

  Filling the NTs changed its $A_\text{free}$, both because the relaxed force-field structure is slightly different than the DFT structure and due to the adsorbed water.
  In Table~\ref{tab:area_rmsd} we report $A^\text{filled}_\text{free}$, the accessible cross-section area computed for the filled NTs configurations, after the removal of all the water molecules inside the NT.
  We see that, overall, the accessible area increased due to the presence of water, which is reasonable considering the NTs are hydrophobic.
  For the (h) (\num{12},\num{0}) NT we observed a small contraction in the presence of water.
  Since this NT only allows a single line of water molecules to be adsorbed (as shown in Figure~\ref{fig:full_isotherms}), the water behaves like it is frozen, which may be related to the smaller $A^\text{filled}_\text{free}$.

  The adsorption isotherms for the selected NTs are presented on the left panel of Figure~\ref{fig:full_isotherms}, reported as the number of adsorbed molecules per NT length to enable the comparison between different NTs.
  We see that the condensation transitions vary between \SI{200}{\mega\pascal} and \SI{750}{\mega\pascal} depending on the NT.
  Comparing these pressures with the $A^\text{filled}_\text{free}$ from Table~\ref{tab:area_rmsd}, we observe a trend indicating that smaller pores (smaller $A^\text{filled}_\text{free}$) lead to a condensation at higher pressure.
  The (m) NT seems to be an exception to this rule, however, when looking at how water is adsorbed in this NT, on the central panel of Figure~\ref{fig:full_isotherms}, we see that there are two separate channels in which water can flow.
  Considering each channel to correspond to half of the $A^\text{filled}_\text{free}$, the effective cross-section accessible area for (m) would be \num{6.49}.
  With this effective $A^\text{filled}_\text{free}$, the condensation transition would be placed between the (c) and (h) NTs, which corresponds to the observed pressure.
  This trend is well illustrated by the cross-section area against the condensation pressure plot shown on the right panel of Figure~\ref{fig:full_isotherms}.

  Comparing the adsorption isotherms for silicalite-1 and the NTs we see that the condensation pressure of the (\num{10},\num{0}) NT (~\SI{220}{\mega\pascal}) is slightly larger than the pressure for silicalite-1 (~\SI{160}{\mega\pascal}).
  Considering the channels of silicalite, which are oblate cylinders formed by 10-membered \ce{SiO2} rings, the two condensation pressures could be expected to be more similar.
  However, the intersection between the two channels effectively increases the accessible volume in comparison with the (\num{10},\num{0}) NT, due to its larger accessible volume.
  Indeed, the intersections of silicalite-1 are known to be able to adsorb water with up to three times the density of the channels.\cite{Desbiens2005}
  According to the $A_\text{free}$ model, this increased accessible volume would lower the condensation pressure, as observed.

  \section{Conclusions}
  \label{sec:conclusions}
  In this work, we have investigated the structure and water adsorption in ultrathin silica NTs.
  Performing annealing with MD simulations and DFT optimizations we found putative global minimum structures for the (\num{10},\num{0}), (\num{12},\num{0}) and (\num{15},\num{0}) \ce{SiO2} NTs.
  As previously reported, the shape of the (\num{12},\num{0}) and (\num{15},\num{0}) is not perfectly cylindrical.
  We measured the accessible cross-section area ($A_\text{free}$) of the NTs and found that the shape reduces $A_\text{free}$ such that it is smaller than the value for the (\num{10},\num{0}) NT.

  Moreover, performing GCMC simulations, we obtained the water adsorption isotherms in silicalite-1 and \num{5} different obtained NTs.
  Comparing with silicalite-1 experimental results, we considered isotherms obtained with three different force fields, selecting the Interface potential to describe the interfacial interactions in the NTs.
  It is worth mentioning that our goal was not to reproduce the silicalite-1 isotherms, but, instead, check which general-purpose silica force-field provided the best isotherm in comparison with the experiment.

  The presence of water overall expanded the NTs, as pure silica without silanol defects is known to be hydrophobic, resulting in larger $A_\text{free}$ after adsorption.
  The obtained condensation pressures varied between \SI{200}{\mega\pascal} and \SI{750}{\mega\pascal} depending on the NT.
  We were able to correlate the condensation pressures with $A_\text{free}$, showing that controlling the NT shape can be important to control the water adsorption.

  \section*{Acknowledgments}
  The authors thank the support of the S\~ao Paulo Research Foundation (FAPESP) grants \#2019/21430-0 and \#2017/02317-2 who financed this work. This work was also partially supported by the Research Council of Norway through the Centre of Excellence \textit{Hylleraas Centre for Quantum Molecular Sciences}  (grant number 262695).

  \section*{Supporting Information}
  The Supporting Information of this work can be found online.

%%%%%%%%%%%%%%%%%%%%%%%%%%%%%%%%%%%%%%%%%%%%%%%%%%%%%%%%%%%%%%%%%%%%%%%%%%%%%%%%%
% BIBLIOGRAPHY

% \clearpage
  \bibliography{references} % Produces the bibliography via BibTeX.
% \clearpage

\end{document}